# An Equation of State for Helium

A Kerley Technical Services Research Report

December 2004

# An Equation of State for Helium

A Kerley Technical Services Research Report

Gerald I. Kerley

December 2004

## ABSTRACT

This report describes a new equation of state (EOS) for helium. The PAN-
DA code was used to construct separate EOS tables for the solid and fluid
phases. The solid and fluid EOS were then assembled into a multiphase
EOS table using the PANDA phase transition option. Contributions from
thermal electronic excitation and ionization were also included in the mod-
el. The EOS gives very good agreement with all static and shock-wave da-
ta, except at temperatures below 20K, where further work is needed. This
EOS was developed primarily for use in models of the giant planets, the
sun, and stars but should be useful in other applications as well.

# CONTENTS

# FIGURES

# TABLES

# 1. INTRODUCTION

Helium is the second most abundant element in the universe and an important constituent of the sun, stars, and giant planets like Jupiter and Saturn. The helium equation of state (EOS) is needed for modeling these heavenly bodies. Though much less abundant on earth, helium is also of scientific interest for testing quantum statistical mechanical theories of matter. In particular, helium undergoes a transition to a superfluid state at low temperatures, and it is the only element that remains in the liquid state at T = 0K.

This report discusses a new EOS table for helium that was generated primarily for use in modeling the giant planets, sun, and stars. It was constructed with the PANDA code [1], using a model similar to that recently applied to hydrogen and deuterium [2]. Reference [3] describes a study of the planets Jupiter and Saturn that makes use of this EOS, along with that for $H_2$ and other compounds.

In constructing this EOS, I have attempted to match all of the experimental data over as wide a range of densities and temperatures as possible. Unfortunately, I was not able to obtain satisfactory results at very low temperatures, T < 20K. An improved treatment of the quantum effects, which dominate the EOS in this regime, would probably remedy this problem. An alternative would be to replace the low-temperature region of the table with a fit to the experimental data.

The new helium EOS table covers the density range $0 \leq \rho \leq 100$ g/cm$^3$ and the temperature range $1 \leq T \leq 1.0 \times 10^8$ K. It is accurate over the entire range tabulated, except for T < 20K, as noted above.

Section 2 of this report gives an overview of the theoretical model and the parameters used in generating the table. Section 3 compares the predictions with experimental data. Conclusions are given in Sec. 4.

## 2. MODEL OVERVIEW

The EOS table for helium was constructed using standard options in the PANDA code [1]. Separate EOS tables were constructed for the solid and fluid phases, and the phase transition option was used to construct a single table including melting and vaporization. This approach has been validated in calculations for a wide variety of materials, most recently for $H_2$ and $D_2$ [2]. The model has already been well documented, and I will only give a brief overview of it here. Details can be found in Refs. [1], [2], and [4].

### 2.1 Solid Phase

The thermodynamic functions for the solid phase were expressed as sums of terms that are assumed to be separable and additive:

$$P(\rho, T) \ = \ P_c(\rho) + P_l(\rho, T) + P_e(\rho, T) \,, \tag{1}$$

$$E(\rho, T) \ = \ E_c(\rho) + E_l(\rho, T) + E_e(\rho, T) - \Delta E_b \tag{2}$$

$$A(\rho, T) \ = \ E_c(\rho) + A_l(\rho, T) + A_e(\rho, T) - \Delta E_b \,. \tag{3}$$

The subscripts $c$, $l$, and $e$ denote contributions from the zero-Kelvin curve, lattice vibrations, and thermal electronic excitations, respectively. These three terms are discussed below. The constant $\Delta E_b$ was chosen to give zero enthalpy for the gas at room temperature and pressure (RTP).

### 2.2 Fluid Phase

The thermodynamic functions for the fluid phase—the liquid, gas, and supercritical regions—were calculated using the PANDA liquid model. The Helmholtz free energy has the form

$$A(\rho, T) \ = \ A_\phi(\rho, T) + A_e(\rho, T) - \Delta E_b \,. \tag{4}$$

Here $A_\phi$ includes the contributions from both the intermolecular forces and the thermal motions of the molecular centers of mass. $A_e$ is the contribution from thermal electronic excitations, the same as in the solid phase. The constant $\Delta E_b$ is also the same as for the solid phase. The other thermodynamic quantities were computed from the usual thermodynamic relations.

The first term in Eq. (4), $A_\phi$, corresponds to the terms $E_c$ and $A_l$ in the solid EOS. It was calculated using a version of liquid perturbation theory called the CRIS model [5][6]. Let $\phi$ be the potential energy of a molecule in the field of neighboring molecules. The free energy can be written in terms of this function by using a perturbation expansion about the properties of an idealized hard-sphere fluid,

$$A_\phi(\rho, T) = A_0(\rho, T, \sigma) + N_0 \langle \phi \rangle_0 + \Delta A_{ho} + \Delta A_{qm}, \tag{5}$$

where $N_0$ is Avogadro's number. Here $A_0$ is the free energy for a fluid of hard spheres, the first-order term $N_0 \langle \phi \rangle_0$ is an average of $\phi$ over all configurations of the hard sphere fluid, $\Delta A_{ho}$ includes all higher-order terms in the perturbation expansion, and $\Delta A_{qm}$ is the quantum correction. The hard-sphere diameter $\sigma$ is defined by a variational principle that minimizes $\left| \Delta A_{ho} \right|$, which selects the hard-sphere system having a structure that is closest to that of the real fluid. The corrections $\Delta A_{ho}$ are then computed from approximate expressions.

It is not currently possible to determine the function $\phi$, which depends upon the intermolecular forces, from either experiment or theory. In the CRIS model, this function is estimated from the zero-Kelvin energy of the solid phase by

$$N_0 \phi \approx (\rho / \rho_s) E_c(\rho_s), \tag{6}$$

where $\rho$ is the actual density of the fluid, and $\rho_s$ is the solid density having the same nearest neighbor distance as that of the given fluid configuration. Equation (6) is then averaged over all nearest neighbor distances using equations derived from the hard-sphere distribution function [5][6]. This approximation has been found to give good results for all kinds of liquids, but it is especially accurate for rare gases and molecular fluids. Hence it is entirely adequate for helium.

In this work, the quantum-mechanical term $\Delta A_{qm}$ was computed using the model developed for the $H_2$ and $D_2$ EOS. As in $H_2$ and $D_2$, the magnitude of this term was controlled by an adjustable parameter $q_E$, chosen to give the best agreement with experimental data. (See Sec. 6.3 of Ref. [2] for further details.) The result is listed in Table 1.

## 2.3 Zero-Kelvin Curve

The zero-Kelvin curve appears in both the solid and the liquid EOS. At low pressures, it was represented by the so-called EXP-N formula in PANDA (Sec. 3.4 of Ref. [1]). The pressure and energy are given by

$$P_c(\rho) = [\rho_0 E_B \nu / (1 - 3\nu/\alpha)] \{ \eta^{2/3} \exp([\alpha(1 - \eta^{-1/3})] - \eta^{\nu+1}) \}, \tag{7}$$

$$E_c(\rho) = [E_B / (1 - 3\nu/\alpha)] \{ (3\nu/\alpha) \exp[\alpha(1 - \eta^{-1/3})] - \eta^\nu \}, \tag{8}$$

where $\eta = \rho/\rho_0$, $\rho_0$ and $E_B$ are the density and binding energy at zero pressure, and $\alpha$ and $\nu$ are constants. The exponential terms in Eqs. (7) and (8) represent the contributions from repulsive forces, while the terms involving $\eta^\nu$ represent the contributions from attractive forces. It is convenient to eliminate the parameter $\alpha$ by relating it to the bulk modulus at zero pressure,

$$\beta_0 = \rho_0 E_B \nu (\alpha/3 - 1/3 - \nu)/(1 - 3\nu/\alpha). \tag{9}$$

The EXP-N formula does not give satisfactory results at high pressures because it does not have the correct asymptotic form as $\rho \to \infty$. To remedy this problem, the PANDA code offers a formula, called the "TFD match", that is used at densities greater than a user-specified value $\rho_m$. (See Sec. 3.6 of Ref. [1] for details.)

In this work, the parameters $E_B$, $\rho_0$, $\beta_0$, and $\rho_m$ were chosen by requiring the model to match static compression data for both liquid and solid helium (after including the other terms in the EOS). For non-polar molecules like helium, the appropriate value of $\nu$ is 2.0. The five zero-Kelvin parameters are listed in Table 1.

**Table 1: Parameters for helium EOS model.**

| parameter | value |
|---|---|
| $E_B$ (MJ/kg) | 0.190 |
| $\rho_0$ (g/cm$^3$) | 0.378 |
| $\beta_0$ (GPa) | 0.490 |
| $\nu$ | 2.0 |
| $\rho_m$ (g/cm$^3$) | 1.0 |
| $\Theta_0$ (K) | 120 |
| $\Gamma_0$ | 2.13 |
| $\tau$ | 0.775 |
| $q_E$ | 1.40 |

## 2.4 Lattice-Dynamical Terms

The lattice-vibrational terms were computed using the Debye model. (The equations for the thermodynamic functions are given in Sec. 4.2 of the PANDA manual.) The following formulas (option IGRN=4 in PANDA) were used to treat the density-dependence of the Debye temperature $\Theta$ and Grüneisen function $\Gamma$.

$$\Theta(\rho) \,=\, \Theta_0(\rho/\rho_0)^{1/2}\exp\{[\Gamma_0 - \Gamma(\rho)]/\tau\}\,, \tag{10}$$

$$\Gamma(\rho) \,=\, (\Gamma_0 - \tfrac{1}{2})(\rho_0/\rho)^{\tau} + \tfrac{1}{2}\,, \tag{11}$$

where $\Theta_0$ , $\Gamma_0$ , and $\tau$ are constants. In the present work, these three parameters were chosen primarily to match the melting curve. However, the values so obtained were also found to give satisfactory agreement with solid EOS data. The three lattice-dynamical parameters are given in Table 1.

## 2.5 Thermal Electronic Terms

The contributions from thermal electronic excitation and ionization to the EOS, subscripted *e* in Eqs. (1)-(4), were computed using the PANDA ionization equilibrium (IEQ) model (Sec. 9 of the PANDA manual). The version used here includes two improvements to the original model that are discussed in a report on the carbon EOS [4]. Additional discussion is given in the $H_2/D_2$ EOS report [2].

The PANDA IEQ model uses the average atom approximation, in that the properties of the system are computed by considering the electronic structure of a single atom in an average "ion sphere" (determined by the density). However, the model explicitly sums over all electronic configurations of the neutral atom and all states of ionization instead of considering a single average configuration.

The starting point for the IEQ model is a table of energies and radii for the one-electron orbitals of the isolated atom in its ground state configuration, including both occupied and unoccupied states. (The "orbdat" file, which is part of the PANDA code package, provides these data for all elements with atomic numbers from 1 to 103 [7].) PANDA uses the orbital data, together with a scaling model, to to generate the energies and statistical weights of all atomic and ionic configurations, along with corrections for continuum lowering and pressure ionization.

The average atom model assumes that charge neutrality holds within an ion sphere. This assumption is reasonable at high densities but is inaccurate at low densities, where the corrections can be computed exactly. The model used here includes corrections for charge fluctuations. It employs an adjustable parameter that controls the interpolation between the low- and high-density limits. (See parameter F3 in Sec. 4.3 of Ref [4].) In the present work, I set F3 = 0.1, the same value that was used for $H_2$, $D_2$, and carbon.

The average atom model also assumes that all ion spheres are equal in volume. In fact, thermal motions can lead to fluctuations in volume. The model used here includes corrections for thermal fluctuations. In practice, the principal effect of these corrections is to smooth discontinuities arising from the cutoff in bound states at pressure ionization. The amount of smoothing is controlled by an adjustable parameter. (See parameter XB in Sec. 4.4 of Ref [4].) In the present work, I initially used XB = $1.0 \times 10^3$, the same as for $H_2$ and $D_2$. However, this value did not give completely satisfactory results, and I had to apply the correction a second time in order to get sufficient smoothing.

The thermal electronic contributions to the entropy and pressure of helium are shown in Fig. 1. The curves show isotherms from 1000 to $1.0 \times 10^7$ K, equally spaced in the logarithm. A striking feature of the plots is the insulator-metal transition that occurs in the density range 10-20 g/cm$^3$. At low densities, the 1s valence electrons are localized and insulating; at high densities, they are pressure ionized and metallic. The nature of the transition is characteristic of elements with closed valence shells and is significantly different from that seen in hydrogen, carbon, and other elements with open shells.

Figure 1 shows that the entropy decreases with density over most of the range depicted—behavior usually seen in both insulators and metals. However, a reversal of this trend is observed at low temperatures, starting with densities above ρ ~1 g/cm$^3$. This effect is caused by continuum lowering—reduction of the 1s ionization energy with increasing density—which increases the degree of ionization. The entropy then increases with density, reaching a maximum value at the transition. At still higher densities, the entropy is once again a decreasing function of density, as expected for a metal.

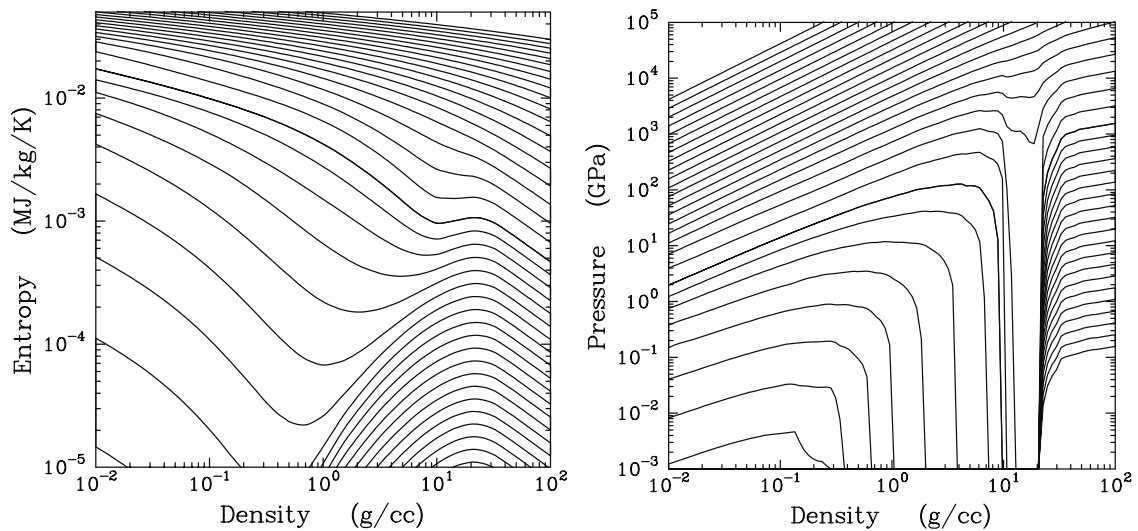

**Fig. 1. Thermal electronic contributions to entropy and pressure for helium. Curves, calculated from the PANDA IEQ model, show 39 isotherms from 1000 to 1.0x10$^7$K, equally spaced in the logarithm.**

It is easy to show, using thermodynamic arguments, that an increase in entropy with density must be accompanied by a *negative* contribution to the pressure, just as seen in Fig. 1. This behavior also occurs in the other rare gases and some elements with closed valence subshells, like the alkaline earth metals. In fact, this phenomenon has a significant effect on the Hugoniot of xenon, where it has been observed experimentally [8]. In helium, however, the transition occurs at a high density and pressure ($\sim 10^4$ GPa), where the electronic contribution is small compared to the other terms in the EOS. Therefore, it has only a minor effect on the thermodynamic properties. Moreover, there are no experimental data that probe the region of the metal-insulator transition in helium at the present time.

## 2.6 Multiphase EOS Table

The PANDA phase transition model (MOD TRN option) was used to compute the melting curve and construct the final EOS for each metal. The density range was $0 \leq \rho \leq 100$ g/cm$^3$, the upper limit corresponding to a pressure of 500 TPa (at 0K). The temperature range was $1 \leq T \leq 1.0 \times 10^8$ K. The mesh points were chosen to give good resolution of the important features of the EOS surface. A tension region was included at temperatures below the boiling point, while Maxwell constructions were included at all higher temperatures. As noted in Sec. 1, however, the low temperature region, T < 20K, is not a good representation of the EOS surface.

Material number 5764 was assigned to this EOS table.

# 3. RESULTS AND DISCUSSION

This section compares the model predictions with experimental data.

## 3.1 Zero-Kelvin Isotherm

Zero-Kelvin isotherms for helium are shown in Fig. 2. Triangles show the EOS fit of Driessen, et al. [9], based on their own measurements and several sets of experimental data taken prior to 1986. Circles show the fit of Loubeyre, et al. [10], based on more recent data at high pressures. The curve calculated from the EOS model (including the zero-point contribution), is shown by the solid line. The agreement is completely satisfactory.[1] The squares, computed from Thomas-Fermi-Dirac (TFD) theory, correspond to the high density limit of the EOS model. (See Sec. 2.3.)

The crosses in Fig. 2 were computed by subtracting the zero-point pressure from the fit of Driessen, et al., using the expressions for the Debye temper-ature and Grüneisen function given in their paper. The calculated curve, shown by the dot-dashed line, is also in satisfactory agreement with those data, showing that the lattice-dynamical model of Sec. 2.4 gives a reasonable description of the solid EOS.

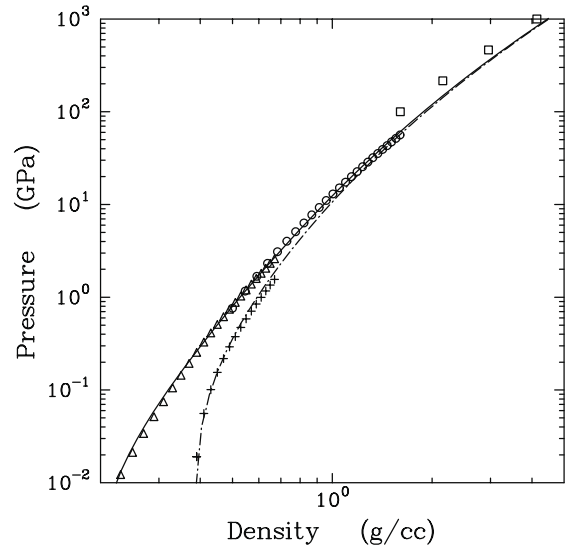

**Fig. 2. Zero-Kelvin curves for helium. Solid and dot-dashed curves from EOS model, with and without zero-point term, respectively. Discrete data points: diamonds and crosses—[9], circles—[10], squares—TFD model.**

## 3.2 Static Compression Data for the Liquid

Figure 3 compares the model predictions with static compression data for liquid helium on several isotherms up to 1000K and pressures up to 2.0 GPa. The triangles are data from the NIST thermophysical data web site [11]. The circles were computed from the fit of Mills, et al. [12].

———————————

1. The calculated curve in Fig. 1 is slightly lower than the fit of Loubeyre, et al., at lower densities and slightly higher at higher densities. However, this small discrepancy is within the error of their fit. In fact, inspection of Fig. 3 in Ref. [10] shows that my curve would give an even better fit to the data than the one given in their paper. Unfortunately, the actual data were only presented in graphical form.

The solid curves, calculated from the EOS model, are in very good agreement with the data for temperatures above 20K. At lower temperatures, the model underestimates the pressure at a given density.

Sensitivity tests using the quantum parameter $q_E$ (Sec. 2.2) show that the quantum mechanical contribution to the pressure is important throughout this regime, even at temperatures as high as 300K. But this term is especially large at low temperatures. This fact suggests that an improved treatment of the quantum-mechanical term could bring the model into better agreement with the experimental data at low temperatures.

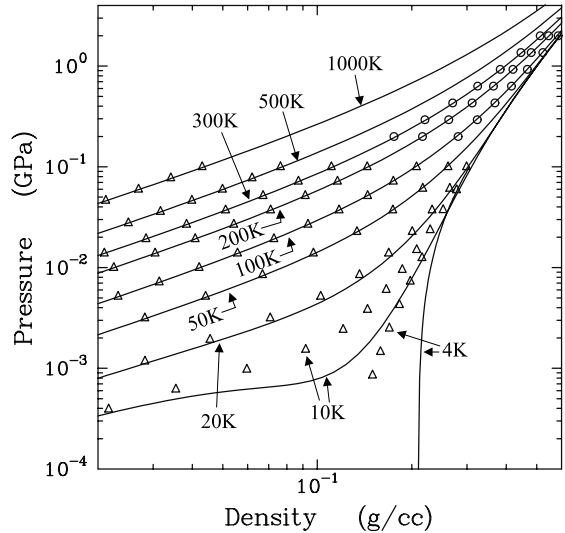

**Fig. 3. Static compression data for liquid helium. Solid lines from EOS model, diamonds from [11], circles from [12].**

## 3.3 Shock Wave Data

Figure 4 compares the model predictions with the four shock wave measurements of Nellis, et al [13]—the only shock data available for helium at present. The three circles are single shock states for an initial density of 0.1233 g/cm$^3$, an initial temperature of 4.30K, and an initial pressure of 0.11 MPa. The square shows a double shock state obtained by reflecting an single shock, with a pressure of 13.8 GPa, off an aluminum anvil.

The solid line in Fig. 4 is the single shock Hugoniot for the new model. It is encouraging to find such good agreement with the measurements, since these data were not used to determine the model parameters.

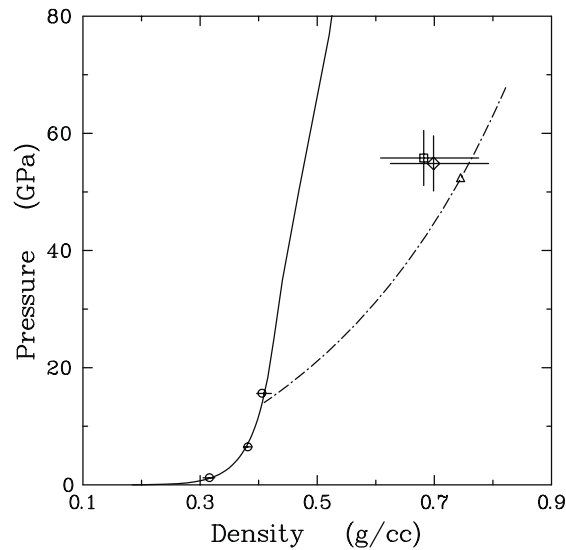

**Fig. 4. Shock wave data for helium. Circles are single shock points, square with error bars is double shock point from [13]. Diamond is revised double shock point, as discussed in text. Solid and dot-dashed curves are model calculations of single and double shock Hugoniots, respectively. Triangle is the result of an impedance-matching calulation, as discussed in text.**

The calculated reflected Hugoniot is shown by a dot-dashed line. In this case, it must be admitted that the agreement with experiment is less satisfying, although the calculated curve does come just within the reported error bars. In fact, my model gives essentially the same result as the one offered by Nellis, et al., along with their data. Their model, like mine, uses liquid perturbation theory. However, they employ an effective pair potential, instead of the zero-Kelvin isotherm, to account for many-body effects. The two approaches have been shown to give comparable results when the intermolecular forces are nearly pairwise additive.

Sensitivity studies show that neither model can give an exact match to both the single shock and the double shock data in Fig. 4. The thermal electronic contributions to the EOS are negligible at the pressures and temperatures of these experiments. Hence, the liquid properties are determined by the intermolecular force law, expressed either as effective pair potential or a zero-Kelvin isotherm. When the intermolecular forces are "stiffened up" to match the double shock point, the single shock Hugoniot becomes too stiff.

In order to be certain that the discrepancy in the double shock point was not due to some problem in analysis of the experiment, I reanalyzed the data given in Table II of Ref. [13], using an accurate EOS for the aluminum anvil [14], correcting for the low initial temperature, and including strength [15]. The result, shown by the diamond in Fig. 4, is only slightly closer to the calculated curve.

It should be noted that the only quantities actually measured in the double shock experiment were the shock velocity in the anvil (8.59 ± 0.17 km/s) and the velocity of the tantalum impactor. (See Ref. [13] for details of the experiment.) Using my theoretical Hugoniots for helium and the anvil, an impedance match calculation gives a shock velocity of 8.479 km/s in the anvil, well within the experimental error bars. The corresponding density-pressure point is shown by the triangle in Fig. 4. I also did a complete numerical simulation of the experiment, using my new helium EOS, together with realistic EOS and constitutive parameters for the anvil and other materials. The shock velocity obtained in the simulation was 8.51 km/s, in good agreement with the impedance match result and the measurement.

Hence it is evident that my new helium EOS is consistent with the double shock point as well as the single shock points. It is obvious that additional shock wave measurements on helium would be very useful for testing this and other EOS models.

## 3.4 Phase Diagram

Figure 5 compares the model predictions of the melting curve with experimental data. Crosses show the fit of Driessen, et al. [9], based on their own measurements and several sets of experimental data taken prior to 1986. Circles show the more recent measurements of Vos, et al. [16]. The calculated curve is shown by the solid line. As noted in Sec. 2.4, the three lattice-dynamical parameters were chosen to give this good fit to the data.

For completeness, Fig. 6 compares the model predictions of the liquid and vapor densities on the coexistence curve with experimental data from Ref. [11]. As already noted, the model does not give satisfactory results at these low temperatures. However, the general shape of the coexistence curve is reasonable. Once again, an improved treatment of the quantum mechanical corrections to the EOS would be expected to bring the model into better agreement with experiment.

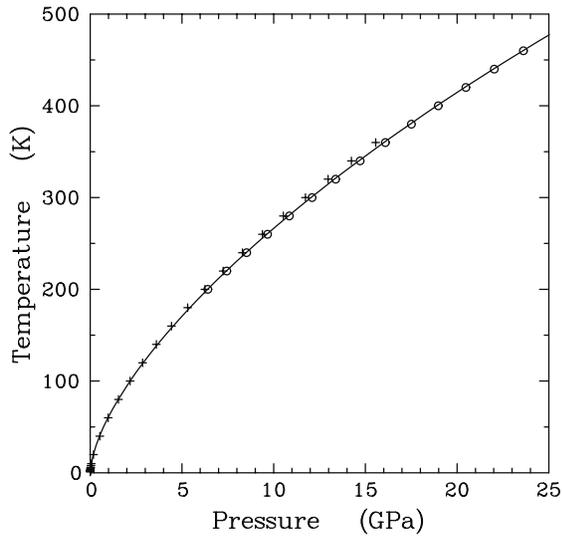

**Fig. 5. Melting curve for helium. Crosses are data from [9], circles are data from [16]. Solid curve is model calculation.**

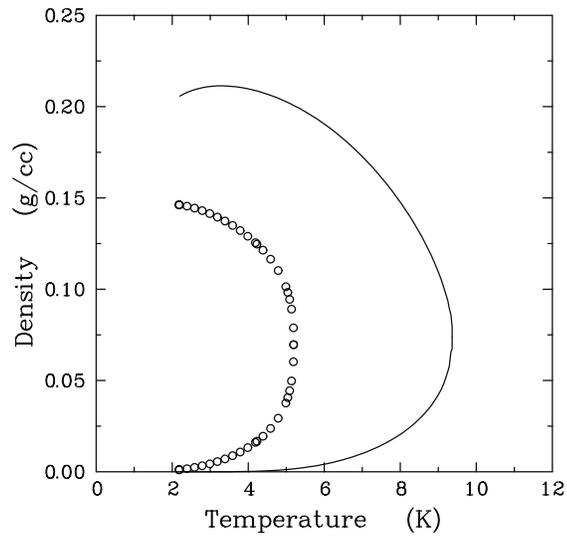

**Fig. 6. Coexistence curve for helium. Circles are data from [11]. Solid curve is model calculation.**

## 4. CONCLUSIONS

This report describes a new EOS for helium that treats all the physical phenomena necessary for use over a very wide range of densities and temperatures—a realistic and accurate treatment of both the solid and fluid phases, the melting curve, and thermal electronic excitation and ionization. It succeeds in matching both the static and shock wave data for helium, using only nine parameters (Table 1).

This EOS was developed primarily for use in modeling the giant planets, sun, and stars. It has already been employed in the study of the planets Jupiter and Saturn, as described in Ref. [3]. Because of its wide range of validity, it should also be useful in other applications.

However, the model is best suited for use at temperatures above 20K. Satisfactory results for the low temperature region will probably require an improved treatment of the quantum mechanical contributions to the EOS in the liquid range. Alternately, it may be possible to replace the tabular data in the low temperature range with an empirical EOS like those given in Refs. [11] and [12].

Additional experimental work on helium, especially more shock wave measurements, would also be very useful.